# Noise-driven informatics: secure classical communications via wire and noise-based computing[1]

**Laszlo B. Kish**

Texas A&M University, Department of Electrical and Computer Engineering, College Station, TX 77843-3128, USA

**Abstract** In this paper, we show recent results indicating that using electrical noise as information carrier offers outstanding potentials reminding of quantum informatics. One example is noise-based computing and logic that shows certain similarities to quantum logic. However, due to the lack of the collapse of wavefunction and due to the immediate accessibility of superposition components, the use of noise-based and quantum computers will probably be different. Another example is secure communications where, out of the unconditional security at idealistic situations, a practical security beyond known quantum solutions can be achieved and has been demonstrated. Here the keys to security are the robustness of classical information, and the second law of thermodynamics. These offer the avoidance of making error statistics and single bit security. It has the potential to restrict the practical applications of quantum communicators to the situations where no wire can be used but optical communication via fiber or via space is possible.

## 1. Introduction

Very recently, it has been shown that thermal noise and its artificial versions (Johnson-like noises) can be utilized as an information carrier [1] with peculiar properties therefore it may be proper to call this topic *Thermal Noise Informatics* [2]. *Thermal Noise Driven Computing*, *Zero Power Communication*, and *Totally Secure Classical Communication* are relevant examples.

---

[1] A short review of earlier results re-edited and expanded, see the references. The noise-based logic results are new and still unpublished at the time of submitting this chapter.

## 2. Zero Power and Zero-Quantum Communications, Stealth Communications

Recently, it has been shown [1] that the equilibrium thermal noise in information channels can be utilized to carry information. In this case, the transmitter does not emit any signal energy into the channel however it only modulates the existing noise there. This issue is completely different from the earlier Porod-Landauer debate [3] about the question if communication without net energy cost is possible by gaining back the energy spent in the communicator devices. (In our opinion, Porod is right, energy-free communication is impossible just like energy-free computing, however those debates are irrelevant here). In our system, the noise is used as information carrier and no effort is made to restore the energy dissipated in the communicator devices. Therefore, this communicator is *not energy-free communication* but *it is free of emitted signal energy*, except perhaps a negligible energy in the order of *kT*/bit, a presently open problem.

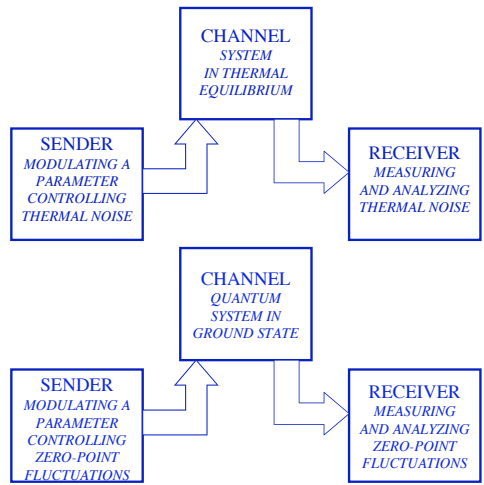

**Fig. 2.1.** Stealth communications. Zero (signal) power classical communication (left) and zero-quantum quantum communication (right) [1].

Zero (signal) power classical communication can utilize the modulation of background thermal noise in the information channel and zero-quantum quantum communication can utilize the modulation of the zero-point fluctuations in the quantum channel, see Figure 2.1, [1].

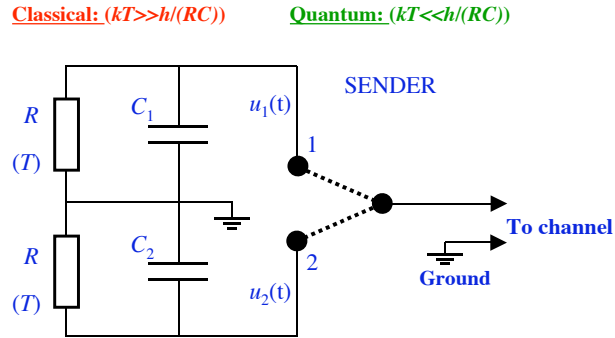

**Fig. 2.2.** A possible realization of stealth communication with zero power classical communication or zero-quantum quantum communication utilizing classical thermal noise and zero-point quantum fluctuations with frequency bandwidth modulation. Classical or quantum, it depends on the upper cutoff frequency $f_c$ and the temperature. Classical: $hf_c \ll kT$; quantum: $hf_c \gg kT$. The receiver can be a simple noise spectrum analyzer or just a simple AC voltmeter.

In [1] some possible realizations were shown. Figure 2.2 shows one of the examples where the Johnson noise of resistors and bandwidth modulation are used in the classical and the quantum limits [1]. In the classical limit, $kT \gg h/RC$, the Johnson noise voltage spectrum is:

$$S_{u,class}(f) = 4kT \, \text{Re}[Z(f)] \, , \tag{2.1}$$

and in the quantum limit, $kT \ll h/RC$, the zero point voltage noise spectrum is:

$$S_{u,quan}(f) = 2hf \, \text{Re}[Z(f)] \, . \tag{2.2}$$

The contacted impedance $Z(f)$ is different in the two positions of the switch because of the different capacitance values. The receiver can be a simple noise spectrum analyzer or just a simple AC voltmeter.

In conclusion, it is possible to execute electronic data communication without injecting signal energy in the information channel. Because this the most invisible way of communication with basically background noise in the channel, it is proper to call these communications *stealth communication*.

## 3. Noise driven computing

Because noise turned out to be a special information carrier with low energy density in the information channel, it is natural to pose the question: can we use it for information processing and computing?

## *3.1. Thermal noise driven computing?*

The first question we addressed is a computer with extraordinarily large error rate, see Fig. 3.1 which we called "thermal noise driven computer". At thermal noise driven computing [4], an idea further inspired by the fact that the neural signals in the brain are noise, certain statistical parameters of an electrical noise, including thermal noise, is the information carrier.

The information may be carried by the bandwidth as at the zero power communication examples above or in other way. To study the minimal energy requirement of requirement/dissipation, another specific realization was used, a simple digital system with zero threshold voltage, when the thermal noise is equal to or greater than the digital signal [4]. Under these conditions, when the digital signal amplitude is less than the variance of the noise, classical digital information is usually considered to be zero or useless, see Figure 3.1.

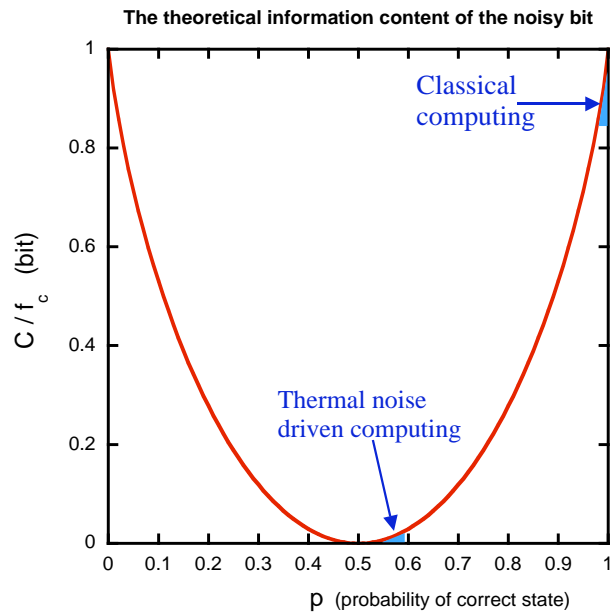

**Fig. 3.1.** Illustration of the range of error probabilities of classical computing and that of the thermal noise driven computing. Y axies: bit content/single clock cycle; plot of Shannon's digital channel coding theorem. A thermal noise driven computer will supposedly work in that range where the information content/ single switching is low, the signal looks like noise and no usual classical computer could work.

We would like to note that in a different class of efforts many notable scientists, including John von Neumann [5]; Forshaw and coworkers [6,7]; and Palem and coworkers [8,9] have been proposing efficient ways of working with probabilistic switches, which are noisy digital logic units with relatively high error probability.

For example, Palem and coworkers have pointed out that this may be a way to reduce power dissipation [8,9] if we give up our requirement of data accuracy. However, though these approaches can be relevant to future developments of thermal noise driven computing, they are very different from our present approach. In a *thermal noise driven computer*, the voltage in the channel is dominantly noise and the modulation of the statistical properties of this noise carry the information. Therefore, the way of extracting the information is not by "*error correcting the information*" like the efforts mentioned above do but by "*decoding the information in the noise*". The realization example we analyzed is working in the regime of huge error probability, $p \approx 0.5$, with zero logic threshold voltage and in the sub-noise signal amplitude limit; and all these parameters are very unusual.

Even though that at this stage there are more open questions than answers and we were are at the moment unable to show a functioning thermal noise driven logical circuitry, certain questions can already be answered about it. Estimations were given about the information channel capacity and the minimal energy dissipation $E_1$ given in *Energy/bit* unit which is given as:

$$\eta = \frac{P_s}{C_{dig}\big|_{p \approx 0.5}} = \frac{\pi \ln 2}{2} kT / bit \approx 1.1 \; bit / kT \tag{3.1}$$

where $P_s$ is the power needed to run the sub-noise level digital signal, $C_{dig}\big|_{p \approx 0.5}$ is he information channel capacity of this signal.

The main advantage of such a hypothetical thermal noise driven computer would be a potentially improved energy efficiency and an obvious lack of leakage current, cross-talk and ground EMI problems due the very low DC voltages. An apparent disadvantage is the large number of extra (redundancy) elements required for error reduction [5-7] if we want a Turing machine.

### *3.2. Noise-based logic: deterministic, with low error rate*

The thermal noise driven computer with the extraordinary error rate shown above is interesting however it is unclear what type of computer it would be. Certainly not a Turing machine because Turing machines must operate error-free and error correction is hopeless here. In this section, we briefly mention a different solution, which excellently fits Turing machines because its error rate can be set arbitrarily low.

In noise-based logic [10] the distinct values are represented by independent stochastic processes: independent voltage (or current) noises. The orthogonality of these processes provides a natural way to construct binary or multi-valued logic circuitry with arbitrary number *N* of logic values by using analog circuitry. Moreover, the logic values on a single wire can be made a (weighted) superposition of the *N* distinct logic values. Fuzzy logic is also naturally represented by a two-component superposition within the binary case (*N*=2). Error propagation and accumulation are suppressed. Other relevant advantages are reduced energy dissipation and leakage current problems, and robustness against circuit noise and background noises such as 1/*f*, Johnson, shot and crosstalk noise. Variability problems are also nonexistent because the logic value is an *AC* signal. A similar logic system can be built with orthogonal sinusoidal signals (different frequency or orthogonal phase) however that has an extra 1/*N* type slowdown compared to the noise-based logic system with increasing number of *N* furthermore it is less robust against time delay effects than the noise-based counterpart.

The noise-based logic we have introduced in [10], in its binary working mode, has the potential to outperform normal digital circuitry even though it is based on analog circuit elements. Still the most far reaching potential of it maybe the high dimensional multi-value logic type working modes provided their mathematical theory is available.

## 4. Unconditionally secure classical communication via wire

### *4.1 The Kirchhoff-loop-Johnson(-like)-noise cipher*

Recently, a totally secure classical communication scheme, a statistical-physical competitor of quantum communicators, was introduced, see Figure 4.1, ([11,12] and [2]), the Kirchhoff-Loop-Johnson(-like)-Noise (KLJN) communicator, with two identical pairs of resistors and corresponding Johnson(-like) noise voltage generators. The KLJN communicators are utilizing the physical properties of an idealized Kirchhoff-loop and the statistical physical properties thermal noise (Johnson noise). The resistors (low bit = small resistor, high bit = large resistor) and their thermal-noise-like voltage generators (thermal noise voltage enhanced by a pre-agreed factor) are randomly chosen and connected at each clock period at the two sides of the wire channel. A secure bit exchange takes place when the bit states at the two line ends are different, which is indicated by an intermediate level of the *rms* noise voltage on the line, or that of the *rms* current noise in the wire. The most attractive properties of the KLJN cipher are related to its security [11,12] and to the extraordinary robustness of classical information when compared that to the fragility of quantum information. To provide security against arbitrary types of attacks, the instantaneous currents and voltages are measured at both ends by Alice and Bob and they are published and compared. In the idealized scheme of the KLJN cipher, the passively observing eavesdropper can extract zero bit (zero-bit security) of information and the actively eavesdropping observer can extract at most one bit while getting discovered (zero-bit security) [11]. The system has a natural zero-bit security against the *man-in-the-middle attack* which is a unique property among secure communicators [13], see Figure 4.2. The KLJN system has recently became network-ready [14]. This new property opens a large scale of practical applications because the KLJN cipher can be installed as a computer card [14], similarly to ethernet network cards. Other practical advantages compared to quantum communicators are the high speed, resistance against dust, vibrations, temperature gradients, and the low price [11]. It has been shown [15] that the KJLN communicator may use currently used wire lines, such as power lines, phone lines, internet wire lines by utilizing proper filtering methods.

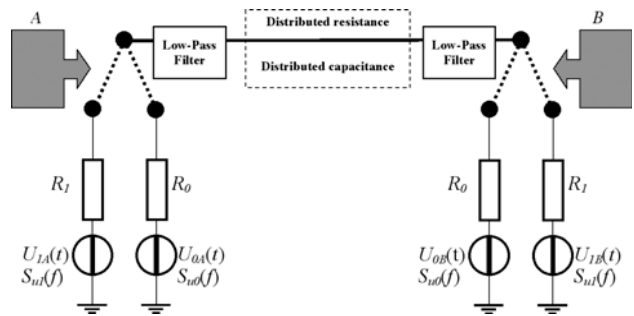

**Fig. 4.1.** The outline of the KLJN communicator. The practical line can generally be represented by distributed resistors and capacitors; however with the proper choice of the cable, the driving elements (resistors and capacitance killer, see below), and the frequency range, it behaves as a simple resistor. The instantaneous current and voltage data are measured and compared at the two ends and, in the case of deviance; the eavesdropping alarm goes on and the currently exchanged bit is not used. The low-pass line filters are necessary to protect against out-of-alarm-frequency-band breaking attempts and false alarms due to parasite transients. False alarms would occur due to any wave effect (transient or propagation effects), illegal frequency components or external disturbance of the current-voltage-balance in the wire.

The fully protected communicator is shown in Fig. 4.3. The fact that classical information is robust and can be monitored continuously provides unconditional security even against any invasive attacks.

Concerning the theoretical security of the KLJN system, the basic rule of any physical secure communication holds here, too: *If you compromise it you lose it*. Though there have been several attempts to break in the KJLN line, so far, no proposed method has been able to challenge the total security of the idealized KLJN system. All these breaking attempts have used various assumptions *directly violating the basic model* of the idealized KLJN method. These assumptions are as follows:

*i.* Allowing non-negligible wire resistance (Bergou [12], Scheuer and Yariv [16]). For the evaluation of this claim at practical conditions, see the response in [17].

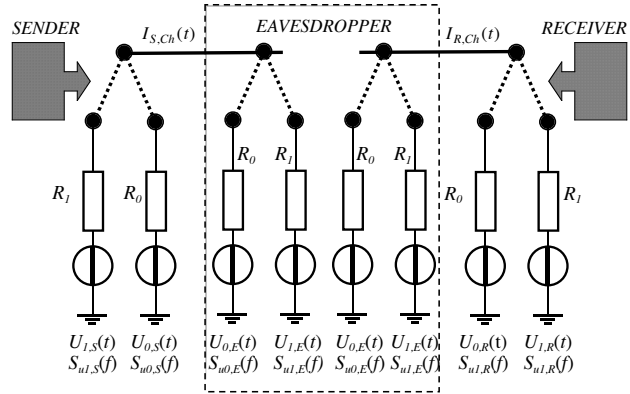

**Fig. 4.2.** An example for the man-at-the-middle-attack by using resistors with the same values and noise voltage generators with the same parameters as those of the sender and the receiver [13]. In the two separate loops the current noises are totally independent thus the attack is immediately discovered within a fraction of the clock duration. Zero bit can be extracted and the alarm goes on. The man-in-the-middle-attack is one of the weakest type of attacks against the KLJN cipher. This property is unique among known communicators.

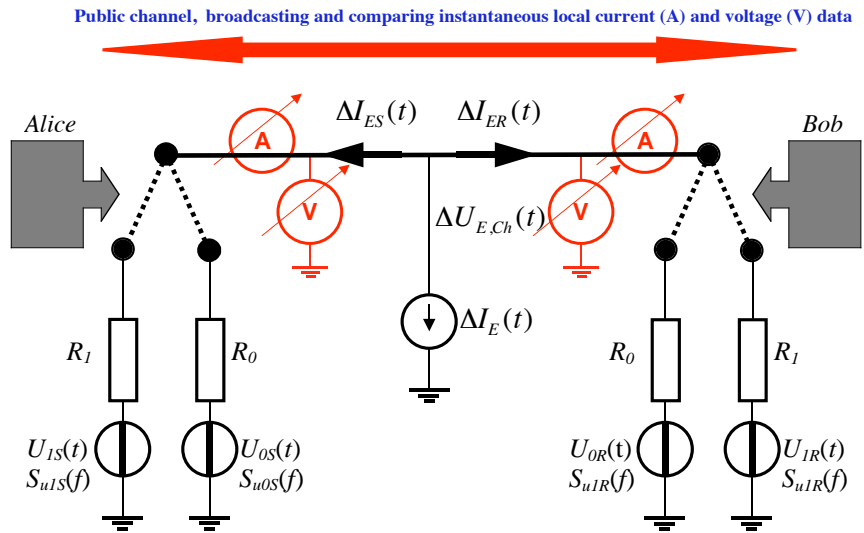

**Fig. 4.3.** The fully protected theoretical KLJN scheme with unconditional security against passive and active eavesdropping [11]. Eve may actively try to extract information by injecting a current at the middle. However, constantly monitoring, broadcasting and comparing the current and voltage values at the two ends uncover the eavesdropping and she can extract no bit without setting on the alarm.

*ii.* Allowing high-enough bandwidth for wave (transient propagation) effects (Scheuer and Yariv [16]). This possibility has been explicitly excluded since the original article [11]. For further refusal of this claim, see the response in [17].

Moreover, *and this is the most direct practical argument*, in this case the standard current and voltage protection of the KLJN system would *immediately alarm and shut down the communication* because the existence of transient and wave effects *per-definition* yield different instantaneous current and/or voltage values at the two ends.

*iii.* Assuming inaccuracy of (noise) temperatures (Hao [18]). For the theoretical refusal of this claim, see [19].

*iv.* Utilizing practical inaccuracy of resistors (Kish [19]). For the mathematical evaluation of this claim at practical conditions, see there [19].

Let us here consider an analogy in the field of quantum communication. Though all the practical quantum communicators suffer from the inability of producing *totally single-photon* output and this fact is unavoidably compromising the security of practical quantum communicators, we do not say that the idealized/mathematical quantum communicator schemes are insecure. In discussing the unconditional security of idealized quantum communication schemes, we suppose that single-photon sources do exist, which is a similar though more difficult claim than to assume the existence of zero wire resistance (c.f. superconductors), and then we conclude that the quantum system is therefore unconditionally secure, at least conceptually. Therefore the practical claim that idealized single-photon source does not exists cannot compromise the claim about the security of an idealized, conceptual quantum communicator. On the other hand, as soon as the security of a *practical communicator system* is discussed, the aspects non-ideal single-photon sources, optical fiber absorption, etc, for quantum communicators and similarly the aspects (*i-iv*) listed above for the KLJN systems cannot be neglected any more and the practical design must keep these effects under control.

The realized KLJN communicator pair shows excellent performance, see Table 4.1, and Fig. 4.4, through a model-line with communication ranges orders of magnitude beyond the ranges of known direct quantum communication channels [21]. The study indicates that the most important type of security leak, which will need most of resources to control, is the Bergou-Scheuer-Yariv type wire resistance effect [12,16,17]; and the rest of the practical security compromises can be neglected, such as the transient/wave arguments of Scheuer-Yariv [16,17], the noise strength argument of Hao [18] and the resistor inaccuracy argument of Kish [19]. The results indicate yet unrivalled fidelity and security levels among existing physical secure communicators and there are straightforward ways to further improve security, fidelity and range, if it is necessary and resources are available, such as thicker cable, etc. Asca comparison, we should keep in mind that the security claims of quantum communicators are usually theoretical because of the expensive nature of relevant breaking tests, while in our practical system we have been able to carry out all the security tests against all the currently known ways of breaking attempts.

| Type of breaking | Measured number, or ratio, of eavesdroppable bits without setting on the current-voltage alarm (tested through 74497 bits) | Remarks |
|---|---|---|
| BSchY (*i*) [12,16] attack in the present KLJN system | 0.19% | 0.00000019% at 10 times thicker wire (theoretical extrapolation). Arbitrarily can be enhanced by privacy amplification [20,21]; the price is slowing down. |
| Hao (*iii*) [16] attack in the present KLJN system | Zero bit | Below the statistical inaccuracy. Considering the 12 bit effective resolution of noise generation accuracy, it is theoretically: $<10^{-10}$ |
| Kish (*iv*) [17] attack utilizing resistor inaccuracies in the present KLJN system | Zero bit | Below statistical inaccuracy. Theoretically, when pessimistically supposing 1% resistance inaccuracy: <0.01% |
| Current pulse injection (Kish) [9] in the present KLJN system. | Zero bit | One bit can be extracted while the alarm goes on thus the bit cannot be used. |

**Table 4.2.** Practical security of the built KLJN cipher [20].

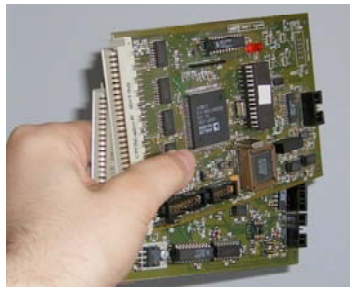

**Fig. 4.4.** The realized KLJN network unit with a pair of KLJN communicators [7]. The computer chip version can be much simpler because of the very short cable distances in the computer providing idealistic conditions to approach the theoretical limits of performance.

*4.2 Unconditionally secure computers and other hardware*

In the case of the need of extraordinary security, Kirchhoff-loop-Johnson-(like)-noise ciphers can easily be integrated on existing types of digital chips in order to provide secure data communication between hardware processors, memory chips, hard disks and other units within a computer or other data processor system, see Fig. 4.5 [22]. This looks like a hopeless goal to do with today's bulky quantum communicators however it is a trivial task for silicon chip technology. The secure key exchange can take place at the very first run and the system can renew the key later at random times with an authenticated fashion to prohibit man-in-the-middle attack. The key can be stored in flash memories within the communicating chip units at hidden random addresses among other random bits that are continuously generated by the secure line but are never actually used. Thus, even if the system is disassembled, and the eavesdropper can have direct access to the communication lines between the units, or even if she is trying to use a man-in-the-middle attack, no information can be extracted. The only way to break the code is to learn the chip structure, to understand the machine code program and to read out the information during running by accessing the proper internal ports of the working chips. However such an attack needs extraordinary resources and even that can be prohibited by a password lockout. The unconditional security of commercial algorithms against piracy can be provided in a similar way.

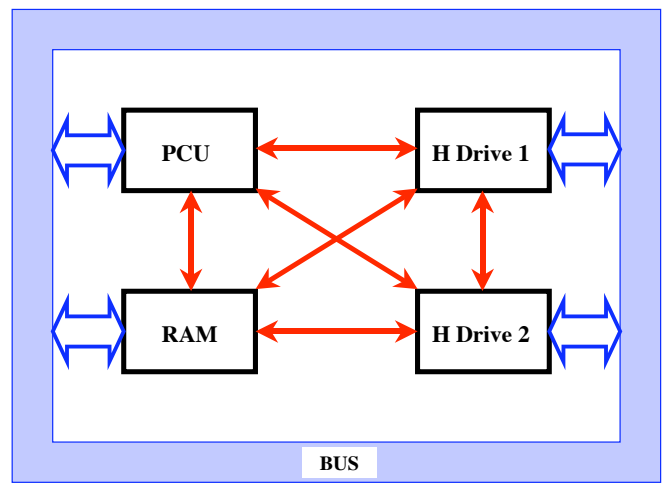

**Fig. 4.5.** Example for securing a subsystem of a PCU, a RAM and two hard drives. Solid arrows: *KLJN* connections; Block arrows: classical data bus connections. Each unit has 3 *KLJN* communicators integrated on their chip. The required number of KLJN units/chip is *N*-1 when *N* chip gets connected securely.

## Acknowledgements

Many colleagues had constructive contributions toward debates shaping the ideas and clarifying the situation with the KLJN cypher. Most of all, Zoltan Gingl and Robert Mingesz, who designed, built and tested the very first KLJN communicator pairs. Among many others, discussions with Terry Bollinger, Oivier Saidi and Bruce Schneier were useful.